\documentclass[review,12pt]{elsarticle}
\usepackage{lineno,hyperref}
\modulolinenumbers[5]

\usepackage{amsmath}
\usepackage{graphicx}

\title{On the amplification of acoustic phonons in carbon nanotube }

\author[rvt]{K. A. Dompreh\corref{cor1}\fnref{fn1}}
\author[focal]{N. G. Mensah}
\author[rvt]{D. Sakyi-Arthur}
\author[rvt]{S. Y. Mensah}
\address[rvt]{Department of Physics, College of Agriculture and Natural Sciences, U.C.C, Ghana.}
\address[focal]{Department of Mathematics, College of Agriculture and Natural Sciences, U.C.C, Ghana}
\cortext[cor1]{Corresponding author\\ Email: kwadwo.dompreh@ucc.edu.gh}
\ead[url]{kwadwo.dompreh@ucc.edu.gh}

\date{}

\begin{document}
\begin{abstract}
\noindent We present a theoretical study of acoustic phonons amplification  
in Carbon Nanotubes (CNT).  The phenomenon is via Cerenkov emission (CE) of 
acoustic phonons using intraband transitions proposed by Mensah et. al.,~\cite{1}
in Semiconductor Superlattices (SSL) and confirmed in ~\cite{2}.
From this, an asymmetric  graph of $\Gamma^{CNT}$  on $\frac{V_d}{V_s}$ 
and $\Omega\tau$ were obtained where amplification ($\Gamma_{amp}^{CNT}$) 
$>>$ absorption ($\Gamma_{abs}^{CNT}$).
The ratio, $\frac{\vert \Gamma_{amp}^{CNT}\vert}{\vert\Gamma_{abs}^{CNT}\vert}\approx 3.5$, 
at $V_d = 1.02V_s$,  $\omega_q = 3.0\ \mathrm{THz}$ and $T = 85\ K$
for scattering angle $\theta > 0$ . A threshold field 
at which $\Gamma_{abs}^{CNT}$ switches over to $\Gamma_{amp}^{CNT}$ was calculated to be 
$E_{z}^{dc} = 6.2\times 10^3\ \mathrm{V/m}$. This
field is far less than  that deduced using Bloch-Type Oscillation (BTO)~\cite{3}
which is $E_{BTO}^{dc} = 3.0\times 10^5\ \mathrm{V/m}$. The obtained $\Gamma_{amp}^{CNT}$
would enable the use of CNT for the production of SASER.\\

\end{abstract}

\maketitle
\section*{Introduction}
The study of amplification of acoustic phonons propagated along the axes of low-dimensional 
and bulk materials such as Semiconductor Superlattices (SL)~\cite{4,5,6,7}, 
$\mathrm{2D}$-Graphene sheet~\cite{8,9}, 
Quantum Wells (QW)~\cite{10,11}, and  Carbon Nanotubes (CNT)~\cite{12,13} are actively persued 
recently using microscopic theory of electron - phonon interactions. 
This is due to the electronic and optoelectronic applications including 
the production of SASER (Sound Amplification by Stimulated Emission of 
Radiation), for dynamic storage of light in quantum wells and acoustic wave
induced carrier transport~\cite{9}.  
When a non-quantizing electric field is applied to these material, and the
drift velocity $V_d > V_s$ ($V_s$ is sound velocity) an amplification of 
acoustic phonons occurs but in the reverse when 
$V_d < V_s$ it lead to absorption of acoustic phonons. 
The nature of this phenomenon
is related to Cerenkov's emission of phonons by moving carriers~\cite{14,15}.
In SSL, amplification of acoustic phonons 
via Cerenkov's emission was proposed theoretically by Mensah et. al.,~\cite{1}. 
The Mensah formalism  of intra-miniband transitions via deformation potential (dp)  
in SSL has been confirmed experimentally by  Shinokita~\cite{2} to amplify  
over $200\%$ of acoustic phonons. Recently, a series of related studies  
in Graphene and CNT with degenerate energy dispersion 
has been conducted (theoretically) in the hypersound regime $ql >> 1$ 
($q$ is the acoustic wavenumber, $l$ is the mean free path). The results obtained qualitatively
agreed with an experimentally obtained results~\cite{12,19}. In multilayer-graphene-based 
system, Yurchenko et. al.,~\cite{16} obtained amplification in the hydrodynamic 
regime $ql << 1$ in a collisionless system with condition $2V_D > V_S$. 
Dagher, et. al.,~\cite{4} utilised the Boltzmann transport equation (BTE) to investigate 
the amplification of travelling waves
in metallic CNT biased by a dc field. The amplification
attained was as a results of Bloch-Type Oscillations (BTO). 
In CNT, the $\pi$-bonding and anti-bonding ($\pi^*$) energy band crosses at the Fermi level
in a linear manner~\cite{17} where, intra-band scattering process depends on the phonon modes. 
These modes are the Longitudinal acoustci (LA), Transverse acoustic (TA) and the
Radial Breathing Mode (RBM) which is the weakest scattering mode. For each phonon branch, an 
electron can be scattered either by zone center phonon, or by a zone 
boundary phonon.  The energy dispersion $\varepsilon(p_z)$ 
is near the Fermi level therefore, at low 
temperatures, conduction occurs through well seperated discrete electron 
states. This leads to the emission of large number of coherent acoustic phonons ~\cite{14,15,18}. 
The extreme electron mobilities  makes CNT, a good candidate for the amplification 
of acoustic phonons. Till date, few studies ~\cite{3,18} has been conducted
to understand amplification in CNT. 
In this paper, we utilised the theory proposed by ~\cite{1} and verified
experimentally by ~\cite{2} to study amplification
of acoustic phonons in CNT. 
The paper is organised as follows: In section $2$, the kinetic theory
based on the linear approximation for the phonon distribution function 
is setup, where, the rate of growth of the 
phonon distribution is deduced  and the amplification is obtained. 
In section $3$,  the final equation is analysed numerically in a graphical 
form  at the harmonic. 
Lastly the  conclusion is presented in section $4$.

\section*{Theory}
We will proceed following the works of ~\cite{18,19,20} where the kinetic equation for the phonon distribution is given as 
\begin{eqnarray}
\frac{\partial N_{{q}}}{\partial t} &=&\frac{2\pi}{\hbar}\sum_p\vert{C_{{q}}}\vert^2 \{[N_{{q}}(t) + 1]f_{{p}}(1-f_{{p}^\prime})
\delta(\varepsilon_{{p}^\prime} - \varepsilon_{{p}} +\hbar\omega_{{p}})\nonumber\\
&-& N_{{q}}(t) f_{{p}^\prime}(1-f_{{p}})\delta(\varepsilon_{{p}^\prime} - \varepsilon_{{p}} +\hbar\omega_{{q}})\}-\gamma N_q(t)\label{Eq_1}
\end{eqnarray}
where $N_{{q}}(t)$ represent the number of phonons with  wave vector ${q}$ at time $t$. The factor $N_{{q}} + 1$ accounts for the 
presence of $N_{{q}}$ phonons in the system when the additional phonon is emitted. The $f_{{p}}(1-f_{{p}})$ represent the probability that 
the initial ${p}$ state is occupied and the final electron state ${p}^\prime$ is empty whilst the factor $ N_{{q}} f_{{p}^\prime}(1-f_{{p}})$
is that of the boson and fermion statistics.  
$\gamma$ denotes phonon losses which includes phonon scattering or phonon
absorption due to non-electronic mechanisms, phonon decay due to anharmonicity of the 
lattice.  In a more convenient form, Eqn. ($1$)
can be written as 
\begin{eqnarray}
\frac{\partial N_{{q}}(t)}{\partial t} = 2\pi\vert{C_{{-q}}}\vert^2[{\frac{N_{{q}}(t) + 1}{1 - exp(\beta(\hbar\omega_{{q}} - \hbar\vec{q}\cdot V_D))}+
{\frac{N_{{q}}}{1 - exp(-\beta(\hbar\omega_{{q}} - \hbar\vec{q}\cdot V_D))}}}]\nonumber\\
\times \sum_{{p}}{(f_{{p}} - f_{{p}^\prime})\delta(\varepsilon_{{p}^\prime} -\varepsilon_{{p}} +\hbar\omega_{{q}})}\label{Eq_3}
\end{eqnarray}
$\beta = 1/k_B T$, $k_B$ is the Boltzmann constant and $T$ is the absolute temperature. 
Here,  phonon loses were ignored and the lowest order in the electron-phonon
coupling is approximated by $f_p$ and 
$$ N_{{q}} = [exp(-\beta(\hbar\omega_{{q}} -\hbar\vec{q}\cdot V_D)-1)]^{-1}$$ 
Eqn.($2$) can further be expressed as 
\begin{eqnarray}
\frac{\partial N_{{q}}(t)}{\partial t} = 2\pi\vert{C_{{q}}}\vert^2[N_{{q}}(t)-\frac{1}{1 - exp(\hbar\omega_{{q}}-\hbar q\cdot V_D)-1}]\nonumber\\
\times ImQ(q,\hbar \omega_q - q\cdot V_D)
\end{eqnarray}
where 
\begin{equation} 
Q =\sum_{{p}}{\frac{f_{{p}} - f_{{p}^\prime}}{\varepsilon_{{p}} -\varepsilon_{p^\prime} 
-\hbar\omega_q -i\delta}}
\end{equation}
and
\begin{equation}
f_{{p}} =[exp(-\beta(\varepsilon_{{p}} -\mu)) + 1 ]^{-1} \label{Eq_5}
\end{equation}
From Eqn. ($3$) the phonon generation rate is given as 
\begin{equation}
\Gamma_{{q}}  = -2\vert{C_{{q}}}\vert^2 Im Q(\hbar\vec{q},\hbar\omega_{{q}} -\hbar\vec{q}\cdot V_D)\label{Eq_6}
\end{equation}
this simplifies to the phonon transition as  
\begin{equation} 
\Gamma_{{q}} = 2\pi\vert{C_{{q}}}\vert^2\sum_{{p}}{(f_{{p}} - f_{{p}^\prime})\delta(\varepsilon_{{p}} 
-\varepsilon_{{p}^\prime} -(\hbar\omega_{{q}} - {\hbar\vec{q}}\cdot V_D))} \label{Eq_7}
\end{equation}
the total rate of absorption and emission of phonon is obtained
by the summation over all the initial and final electrons states.
In Eqn.($7$),  $f_{{p}} > f_{{p}\prime}$ if $ \varepsilon_{{p}} < \varepsilon_{{p}^\prime}$. When 
$\hbar\omega_{{q}} - \hbar\vec{q} \cdot V_D > 0$, the system would return to its equilibrium configuration when perturbed 
but  $\hbar\omega_{{q}} - \hbar\vec{q} \cdot V_D < 0$ leads to the 
Cerenkov condition of phonon instability (amplification).  
From perturbation theory, the transition probability per unit time from 
the initial state $\vert p \rangle$, consisting of electron having 
momentum $p_z$, to the final state $\vert p^{\prime}\rangle$, which consists of an 
electron with momentum $p_z^{\prime}$ and a phonon with wave vector $q$. 
The phonon and the electric field 
are directed along the CNT axis therefore ${p_z}^\prime = ({p}+\hbar {q})cos(\theta)$  where 
$\theta$ is the scattering angle.
In CNT, the linear energy dispersion $\varepsilon({p})$ relation 
is given as~\cite{13}
\begin{equation} 
\varepsilon({p_z}) = \varepsilon_0 \pm \frac{\sqrt{3}}{2\hbar}\gamma_0 b({p_z} - {p}_0) \label{Eq_8}
\end{equation}
The $\varepsilon_0$ is the electron energy in the Brillouin zone at momentum $p_0$, $b$ is the 
lattice constant , $\gamma_0$ is the tight binding 
overlap integral ($\gamma_0 = 2.54$eV). The $\pm$ sign indicates that in the vicinity of the tangent point,
the bands exhibit mirror symmetry with respect to each point. At low temperature, when, $k_B T << 1$,  Eqn.($5$) reduces to
\begin{equation} 
f_{{p}} = exp(-\beta(\varepsilon(p_z)-\mu)) 
\end{equation}
Inserting Eqn.($8$ and $9$) into  Eqn.(7), and after some cumbersome calculations yield
\begin{equation}
\Gamma^{CNT} = \frac{4\hbar\pi\vert C_{{q}}\vert^2exp(-\beta(\varepsilon_0 -\chi {p}_0))}
{\gamma_0 b\sqrt{3}(1-cos(\theta))} \{exp(-\beta\chi(\eta +\hbar {q})cos(\theta))- exp(-\beta\chi\eta)\}\label{Eq_10}
\end{equation}
where $\chi = {\sqrt{3}\gamma_0 b}/{2\hbar}$, and 
\begin{equation}
\eta =\frac{2{\hbar}^2\omega_{{q}}(1 -\frac{V_d}{V_s}) + \gamma_0 b\sqrt{3}\hbar {q} cos(\theta)}{\gamma_0 b\sqrt{3}(1-cos(\theta))}
\end{equation} 

\section*{Analysis}
In the formulation, we utilise a perturbation theory of electron transition 
where, electrons are assumed to drift relative to the lattice ions. 
The wavelength of the phonon is short compared with the screening length 
for the electrons. Electron-electron interactions and phonon loses 
are ignored but the electron-phonon interaction $C_q$ is asumed to be weak 
and treated as perturbation. In Eqn. ($1$), the quantum-mechnical matrix element describing the electron-phonon
coupling for a highly excited (intense acoustic wave) phonon is 
$\vert C_{-q}\vert^2\approx\vert C_q\vert^2$.  
Considering the finite electron concentration, the matrix element can be modified as 
\begin{equation} 
\vert C_{{q}}\vert^2\rightarrow \frac{\vert C_q\vert^2}{\vert{\aleph^{(el)}({q})\vert^2}}
\end{equation}
where $\aleph^{(el)}({q})$ is the electron permitivity. However, for acoustic phonons, 
$\vert{C_{{q}}}\vert = \sqrt{{\Lambda^2 \hbar {q}}/{2\rho V_s}}$,
where $\Lambda $ is the deformation potential constant and $\rho$ is the density of the material.
From Eq.($10$), taking $\varepsilon_0 = {p}_0 = \mu = 0$, the Eqn.($10$) finally reduces to  
\begin{equation}
\Gamma^{CNT} =\frac{\vert {\Lambda}\vert^2\hbar^3 q^2 exp(-\beta\chi\eta)}{2\pi{\hbar\omega_q}\gamma_0 b\sqrt{3}(1-cos(\theta))} 
\{\sum_{n=-\infty}^{\infty}{\frac{exp(-n(\theta)+\beta\chi\eta)}{I_n(\beta\chi(\eta +\hbar {q}))} - 1}\} 
\end{equation}
where $I_n(x)$ is the modified Bessel function.  When the scattering angle $\theta = 0$, $\Gamma = \infty$ whereas
for $\theta > 0$ and  $V_d > V_s$,  $\Gamma^{CNT}$ changes sign from positive ($+$) to ($-$) and amplification is obtained. 
To analyse Eqn. ($15$), the following parameters
were used: $\vert {\Lambda}\vert = 9$eV, $b = 1.42\times 10^{-9}$m,
$q = 10^4$ m$^{-1}$, $\omega_q = 10^{12}$s$^{-1}$, $V_s = 4.7\times10^3$ m s$^{-1}$, $T = 85\ K$, and $\theta > 0$. 
The choice of these parameters espercially that of the acoustic wavenumber 
is based on our previous studies ~\cite{18}.
The dependence of $\Gamma^{CNT}$ on $\frac{V_d}{V_s}$ 
at $n = 1$  is presented below (see Figure $1a$).
\begin{figure}[!h]
\includegraphics[width=7.5cm]{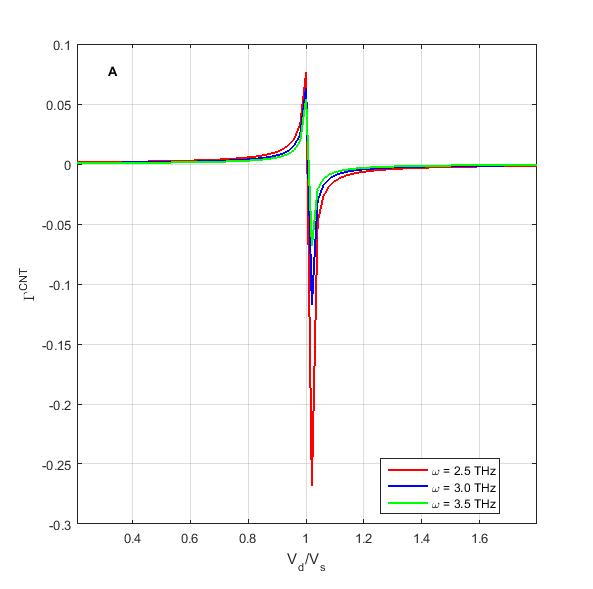}
\includegraphics[width=7.5cm]{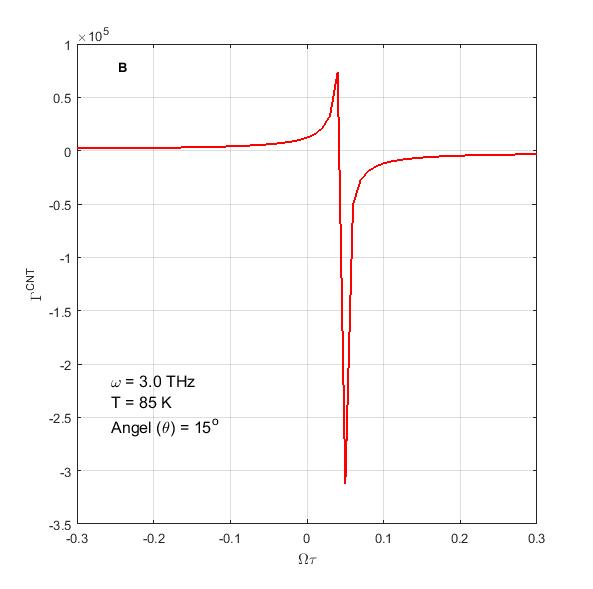}
\caption{Dependance of $\Gamma^{CNT}$ on: (a) $\frac{V_d}{V_s}$ at various 
$\omega_q$ (b) $\Omega\tau$ at $T = 85\ K$, $\omega_q = 3\ THz$ and $\theta = 15^o$}
\end{figure}
From the graph, there is absorption ($\Gamma_{abs}^{CNT}$) when $V_d < V_s$
but when $V_d > V_s$ it swiches over to amplifcation ($\Gamma_{amp}^{CNT}$). 
This satisfy the Cerenkov condition for acoustic phonon emission. The maximum $\Gamma_{amp}^{CNT}$
occurred at $V_d = 1.02V_S$. 
The amplification obtained far exceed the absorption.
The ratio of the amplification to absorption 
\begin{equation}
\frac{\vert \Gamma_{amp}^{CNT}\vert}{\vert\Gamma_{abs}^{CNT}\vert}\approx 3.5
\end{equation}
To determine the threshold field at which $\Gamma_{abs}^{CNT}$ switches  to 
$\Gamma_{amp}^{CNT}$, we calculated and found the $V_d$ to be 
\begin{equation}
V_d = \frac{8\gamma_0}{\sqrt{3}\hbar bm}\sum_{r=1}^{\infty}
\frac{r^2\Omega\tau}{1 + (r\Omega\tau)^2} \sum_{s=1}^m F_{rs}E_{rs}
\end{equation}
Here, 
\begin{equation}
F_{rs} = \frac{a}{2\pi}\int_0^{\frac{2\pi}{a}}
{\frac{exp(-iarp_z)}{1 + exp(\varepsilon(p_z)/k_{\beta}T)} dp_z}
\end{equation}
and 
\begin{equation}
E_{rs} = \frac{a}{2\pi}\int_0^{\frac{2\pi}{a}}
\varepsilon(p_z)exp(-irp_z)dp_z
\end{equation}
where $\Omega = eaE$ ($E$ is the electric field, r the radius 
of the CNT, and $a = 3b/2\hbar$).
The $V_d$ is solved from the 
Boltzmann kinetic equation in the $\tau$-approximation ~\cite{21,22,23}. 
The justification for the $\tau$-approximation can be found in ~\cite{17}. 
By  substituting Eqn.($15$) into Eqn.($11$).  
A graph of $\Gamma^{CNT}$ against $\Omega\tau$
is plotted in Figure ($1b$). 
It can be observed that the threshold field for which $\Gamma_{abs}^{CNT}$ changes over 
to $\Gamma_{amp}^{CNT}$ occurs at $\Omega\tau = 0.04$ which gives $E_z^{dc} = 6.2\times 10^3\ \mathrm{V/m}$.
This value is far lower than that calculated by Dagher et. al.~\cite{4} to be 
$E_z^{dc} = 3\times 10^5\ \mathrm{V/m}$ using 
Bloch-type oscillations (BTO). This is of the order of $2$ which is quite 
high. 

\section*{Conclusion}
We  studied the amplification of acoustic phonon in CNT theoretically in the 
hypersound regime. 
The method used involves the  Cerenkov emission of acoustic phonons
where, when $V_d < V_s$, gave an absorption ($\Gamma_{abs}^{CNT}$),
but when $V_d > V_s$, 
an amplification ($\Gamma_{amp}^{CNT}$) was obtained.
The ratio $\frac{\vert \Gamma_{amp}^{CNT}\vert}{\vert\Gamma_{abs}^{CNT}\vert}\approx 3.5$,
at $V_d = 1.02V_s$, and $\omega = 3.0\ \mathrm {THz}$. 
The threshold field at which  $\Gamma_{abs}^{CNT}$ switched over to $\Gamma_{amp}^{CNT}$
was calculated as $E_z^{dc} = 6.2\times 10^3\ \mathrm{V/m}$. This is far lower 
than that calculated via the BTO to be $E_z^{dc} = 3\times 10^5\ \mathrm{V/m}$.
We therefore propose the use of Carbon Nanotube as a material for the production
of SASER.

\renewcommand\refname{Bibliography}
\section*{Bibliography}

\end{document}